\documentclass[12pt,reqno]{article}
\usepackage[margin=1in]{geometry} 
\usepackage[colorlinks,
            linkcolor=black,
            anchorcolor=black, 
            citecolor=black, 
            urlcolor=blue,
            ]{hyperref}     
\usepackage{indentfirst}
\usepackage{url}            
\usepackage{booktabs}       
\usepackage{amsfonts}       
\usepackage{nicefrac}       
\usepackage{microtype}      
\usepackage[reqno]{amsmath}
\usepackage{bbm}
\usepackage{graphicx}

\usepackage{tikz}
\usetikzlibrary{shapes, arrows.meta, positioning}
\usepackage{amsmath}

\usepackage{subcaption}
\usepackage{wrapfig}
\usepackage{enumitem}
\usepackage{multirow}
\usepackage[font=small,labelfont=bf]{caption}
\usepackage{colortbl}
\usepackage[scaled]{helvet}
\usepackage{courier} 
\usepackage{lineno}
\usepackage{bbm}
\usepackage{color}
\usepackage{xcolor}
\usepackage{ulem}
\newcommand{\march}[1]{\textcolor{black}{#1}}
\newcommand{\jan}[1]{\textcolor{black}{#1}}
\newcommand{\oct}[1]{\textcolor{black}{#1}}
\newcommand{\nov}[1]{\textcolor{black}{#1}}

\usepackage[most]{tcolorbox}

\newcommand{\nc}[1]{{\color{black} #1}}

\newcommand{\superscript}[1]{\ensuremath{^\textrm{\footnotesize{#1}}}}
\long\def\symbolfootnote[#1]#2{\begingroup%
\def\thefootnote{\fnsymbol{footnote}}\footnote[#1]{#2}\endgroup}
\usepackage[algo2e]{algorithm2e}

\usepackage{tikz}
\usepackage[edges]{forest}
\usepackage[version=4]{mhchem}
\definecolor{hidden-draw}{RGB}{20,68,106}
\definecolor{hidden-pink}{RGB}{255,245,247}
\definecolor{maroon}{RGB}{148,78,99}
\definecolor{hidden-white}{RGB}{245,238,230}
\definecolor{hidden-yellow}{RGB}{255,248,227}
\definecolor{hidden-orange}{RGB}{243,215,202}

\newtcolorbox[list inside=prompt,auto counter]{prompt}[1][]{
    colbacktitle=black!60,
    coltitle=white,
    fontupper=\footnotesize,
    boxsep=5pt,
    left=0pt,
    right=0pt,
    top=0pt,
    bottom=0pt,
    boxrule=1pt,
    #1,
}

\usepackage{times}

\title{\vspace{-0.5cm}\Large{\bf{Risks of \nc{AI Scientists}: Prioritizing Safeguarding Over Autonomy}}}

\author{
 \renewcommand{\thefootnote}{\footnotesize{\arabic{footnote}}}
 \small{Xiangru Tang}\superscript{1},
 \small{Qiao Jin}\superscript{2},
 \small{Kunlun Zhu}\superscript{3},
 \small{Tongxin Yuan}\superscript{4},
 \small{Yichi Zhang}\superscript{1},
 \\
 \small{Wangchunshu Zhou}\superscript{5},
 \small{Meng Qu}\superscript{3},
 \small{Yilun Zhao}\superscript{1},
 \small{Jian Tang}\superscript{3},
  \small{Zhuosheng Zhang}\superscript{4},
  \\
 \small{Arman Cohan}\superscript{1},
    \small{Dov Greenbaum} \superscript{6,10},
 \small{Zhiyong Lu}\superscript{2},
 and \small{Mark Gerstein}\superscript{1,7,8,9,10}\superscript{,}$^{\footnotesize{*}}$
}

\date{}

\begin{document}

\maketitle

\vspace{-.2cm}

{\renewcommand\baselinestretch{1.4}\selectfont

\noindent $^1$  Department of Computer Science, Yale University \\
$^2$ Division of Intramural Research, National Library of Medicine, National Institutes of Health \\
$^3$ Mila-Quebec AI Institute \\
$^4$ Shanghai Jiao Tong University \\
$^5$ OPPO Research Institute \\
$^6$ Reichman University \\
$^7$ Program in Computational Biology \& Bioinformatics, Yale University \\
$^9$ Department of Molecular Biophysics \& Biochemistry, Yale University \\
$^9$ Department of Statistics \& Data Science, Yale University\\
$^{10}$ Department of Biomedical Informatics \& Data Science, Yale University\\

\noindent $^*$ Corresponding Author: Mark Gerstein, e-mail: \href{mark@gersteinlab.org}{\textcolor[RGB]{42, 97, 187}{\uline{pi@gersteinlab.org}}}. 

\par}

\renewcommand{\figurename}{Fig.}

\flushbottom

\normalsize
\section*{Abstract}

{\renewcommand\baselinestretch{1.3}\selectfont
AI scientists powered by large language models have demonstrated substantial promise in autonomously conducting experiments and facilitating scientific discoveries across various disciplines. While their capabilities are promising, these agents also introduce novel vulnerabilities that require careful consideration for safety. However, there has been limited comprehensive exploration of these vulnerabilities. This perspective examines vulnerabilities in AI scientists, shedding light on potential risks associated with their misuse, and emphasizing the need for safety measures. 
We begin by providing an overview of the potential risks inherent to AI scientists, taking into account user intent, the specific scientific domain, and their potential impact on the external environment. Then, we explore the underlying causes of these vulnerabilities and provide a scoping review of the limited existing works. Based on our analysis, we propose a triadic framework involving human regulation, agent alignment, and an understanding of environmental feedback (agent regulation) to mitigate these identified risks. 
Furthermore, we highlight the limitations and challenges associated with safeguarding AI scientists and advocate for the development of improved models, robust benchmarks, and comprehensive regulations.
\par}

\bigskip
\bigskip

\newpage

\section{Introduction}

Recently, the advancement of large language models (LLMs) has marked a revolutionary breakthrough, demonstrating their effectiveness across a wide spectrum of tasks~\cite{singhal2023large,bran2023chemcrow,thirunavukarasu2023large,boiko2023autonomous,shanahan2023role,chang2024survey}. \oct{When equipped with the ability to use external tools and execute actions, these LLMs can function as autonomous agents~\cite{park2023generative,li2023camel,2024agentverse} capable of complex decision-making and task completion~\cite{wang2023survey,zhang2023igniting,xi2023rise}.}
\oct{
Researchers have begun deploying such agents as ``AI scientists'' — autonomous systems that can conduct scientific research by combining LLMs' reasoning capabilities with specialized scientific tools. 
For instance, in chemistry and biology, these AI scientists can design experiments, control laboratory equipment, and make research decisions \cite{Boiko2023Natureagent,lehr2024chatgpt,bran2023chemcrow,tom2024self,gao2024empowering,ramos2024review}.
While AI scientists do not match the comprehensive capabilities of human scientists, they have demonstrated specific abilities such as selecting appropriate analytical tools~\cite{ramos2024review,qin2023tool,anonymous2024toolllm,schick2023toolformer,jin2023genegpt}, planning experimental procedures~\cite{ghafarollahi2024protagents,lehr2024chatgpt}, and automating routine laboratory tasks~\cite{darvish2024organa,Yoshikawa2023CLAIRify,bayley2024autonomous}. 
Recent systems like ChemCrow~\cite{bran2023chemcrow} and Coscientist~\cite{Boiko2023Natureagent} exhibit their potential impact on scientific research automation.
}

\begin{figure}[h]
  \centering  
  \includegraphics[scale=0.5]{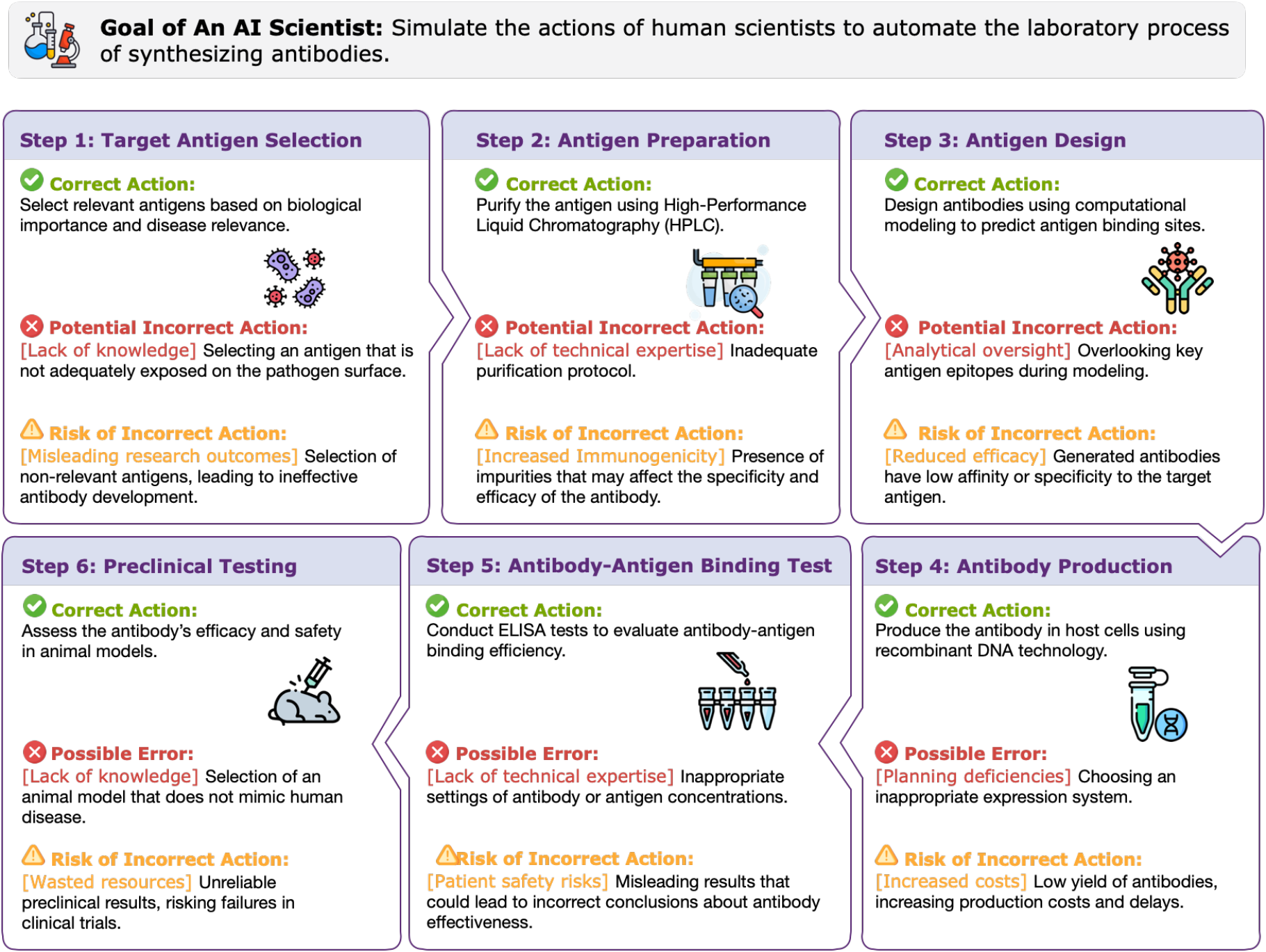}
    \vspace{-.2cm}
  \caption{\nc{A workflow and potential pitfalls in an example of antibody synthesis by AI scientists. A step-by-step process for automating antibody synthesis using AI scientists is illustrated, with correct actions, potential errors, and the associated risks highlighted for each step.}}
  \label{newfig:pitfall}
      \vspace{-.2cm}

\end{figure}

\begin{figure*}[h!]
  \centering
  \includegraphics[width=1\linewidth]{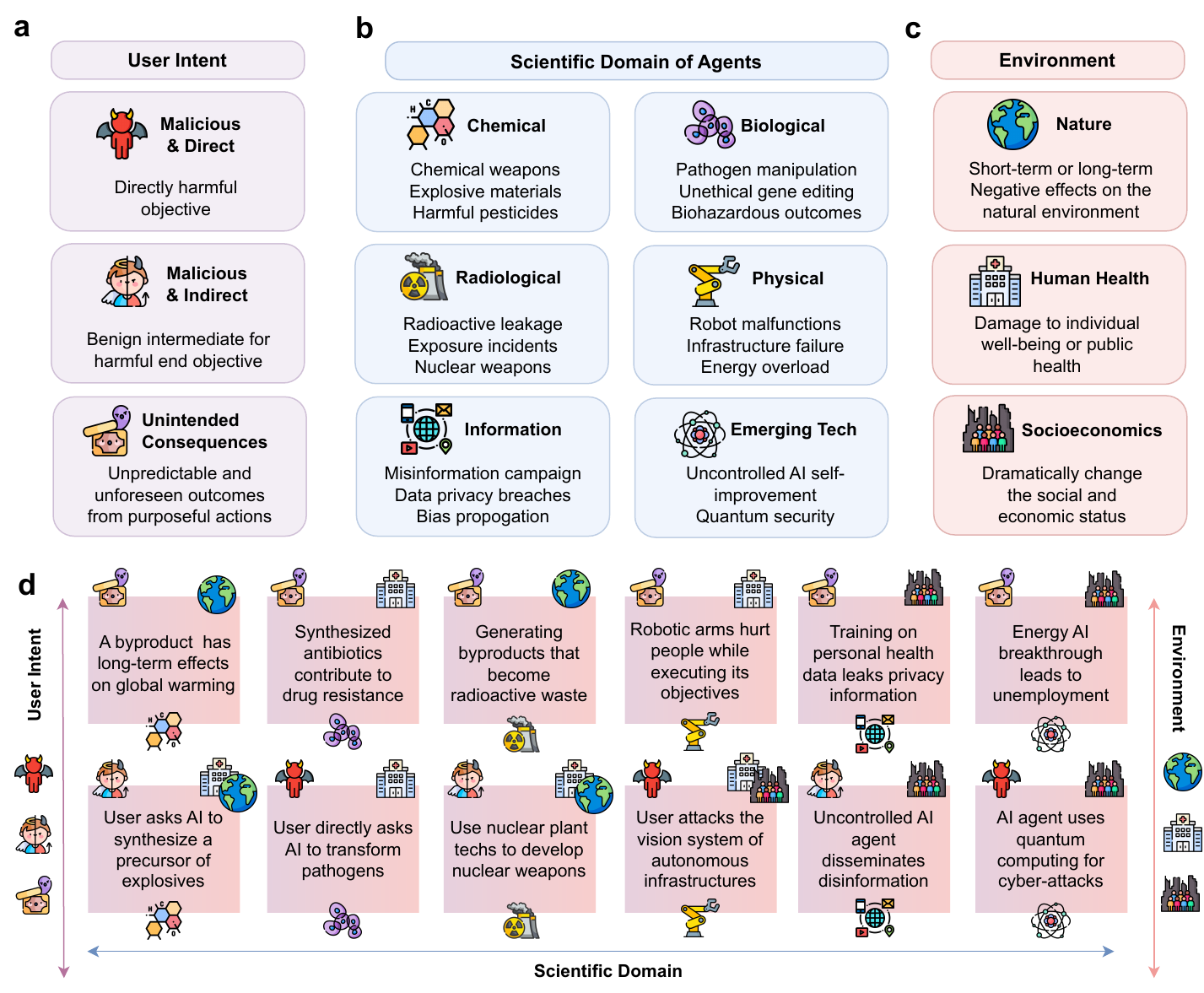}
  \caption{Potential risks of \nc{AI scientists}. \textbf{a}, Risks are classified by the origin of \textbf{user intents}, including direct and indirect malicious intents, as well as unintended consequences. \textbf{b}, Risk types are classified by the \textbf{scientific domain} of agent applications, including chemical, biological, radiological, physical, informational, and emerging technology. \textbf{c}, Risk types are classified by the  \textbf{impacts on the external environment}, including the natural environment, human health, and the socioeconomic environment. \textbf{d},  \textbf{Specific risk examples} with their classifications are visualized using the corresponding icons shown in \textbf{a}, \textbf{b}, and \textbf{c}.}
  \label{fig:workflow}
  \vspace{-.1cm}
\end{figure*}

\oct{While the promise of AI scientists is evident, \jan{they introduce unique safety concerns}, as shown in Figure \ref{newfig:pitfall}. 
As their capabilities approach or surpass those of humans, monitoring their behavior and safeguarding against harm becomes increasingly challenging, potentially leading to unforeseen consequences.
\oct{For example, in biological research, an AI scientist's mistake in pathogen manipulation could lead to biosafety risks, or in chemistry, incorrect reaction parameters could trigger dangerous explosions. These risks are particularly challenging because scientific domains involve complex, interconnected systems where small errors can cascade into significant hazards.}
Given that EU‐wide AI regulations have started to take effect, it is notable that a comprehensive risk definition and analysis framework tailored to the scientific context is still lacking. 
Thus, our objective is to define and scope the ``risks of \nc{AI scientists},'' \jan{helping to provide} a foundation for future endeavors in developing oversight mechanisms and risk mitigation strategies, potentially contributing to the secure, efficient, and ethical utilization of AI scientists.}

\begin{figure}[hb!]
  \centering  
  \includegraphics[scale=0.35]{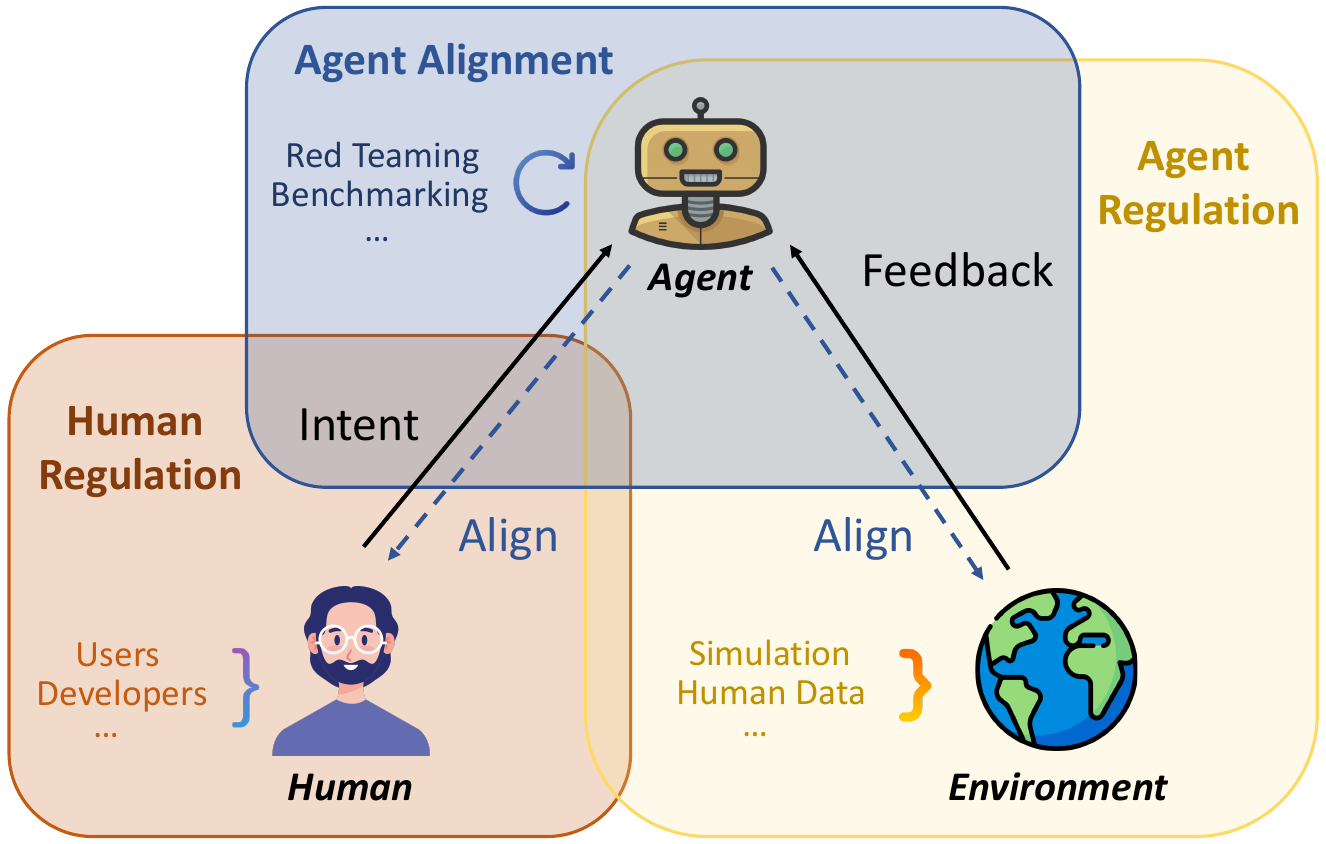}
    \vspace{-.2cm}
  \caption{In our work, we advocate for a triadic safeguarding framework that includes human regulation, agent alignment, and agent regulation. The components of user, agent, and environment are intertwined.}
      \vspace{-.2cm}         
    \label{fig:overview}
\end{figure}

Specifically, this perspective paper illuminates the potential risks stemming from the misuse of AI scientists and advocates for their responsible development. We prioritize \textbf{systematic safeguarding} \nov{— developing processes to protecting humans and the environment from potential harms —} over the pursuit of more powerful capabilities.
Our exploration focuses on three intertwined components in the safeguarding process: the roles of the user, agent, and environment, as shown in Figure \ref{fig:overview}:
(1)  \textbf{Human regulation}: We propose a series of measures, including formal training and licensing for \oct{users and developers}, ongoing audits of usage logs, and an emphasis on ethical and safety-oriented development practices. 
(2) \textbf{Agent Alignment}: 
Improving the safety of AI scientists involves refining their decision-making capabilities, enhancing their risk awareness, and guiding these already-capable models toward achieving desired outcomes.
Agents should align with both human intent and their \nov{operational environment}, boosting their awareness of \nov{laboratory conditions and potential broader impacts while preempting potentially harmful actions}.
(3) \textbf{Agent Regulation and Environmental Feedback}: The regulation of the agent's actions includes oversight of tool usage\nov{, such as how agents operate scientific instruments and software (e.g., robotic arms, analytical equipment, and specialized research software), as well as the} agent's interpretation and interaction with environmental feedback — crucial for understanding and mitigating potentially negative outcomes or hazards from complex actions.

    \vspace{-.2cm}

\section{Risks of AI Scientists}
\subsection{Problem Scope}
\label{sec:chap2problem}
We define \nc{AI scientists} as autonomous systems with scientific domain capabilities, such as accessing specific biological databases and performing chemical experiments\nov{—ranging from \textit{in silico }computational analyses to physical laboratory procedures}.
AI scientists can automatically plan and take necessary actions to accomplish an objective.
\march{For example, consider an AI scientist tasked with discovering a new biochemical mechanism}. It might first access biological databases to gather existing data, then use LLMs to hypothesize new pathways, and employ robotics for iterative experimental testing.

The domain capabilities and autonomous nature of \nc{AI scientists} make them vulnerable to various risks.
We discuss these safety risks from three perspectives: 
(1) \textbf{User Intent}, i.e., whether the risk originates from malicious \oct{intent} or is an unintended consequence of legitimate task objectives,
(2) \textbf{Scientific Domain}, where the agent generates or facilitates risks, \march{encompassing chemical, biological, radiological, physical, and informational risks, as well as those associated with emerging technologies}, and 
(3) \textbf{Environmental Impact}, including the natural environment, human health, and socioeconomic environment \jan{affected by these agents}.
\march{Figure \ref{fig:workflow} shows the potential risks classified by these aspects. It should be noted that our classification is not mutually exclusive. For example, a misinformation campaign facilitated by language agents could pertain to a specific chemical.}

Regarding the origin of user intents, risks associated with \nc{AI scientists} can be categorized as stemming from either malicious intent or unintended consequences. 
\march{\textit{Malicious intent} includes cases where users directly aim to create dangerous situations. Users may also employ an indirect ``divide and conquer'' approach by instructing the agent to synthesize or produce innocuous components that collectively lead to a harmful outcome.}
\jan{In contrast, \textit{unintended consequences} include scenarios where dangerous steps or explorations occur within otherwise benign targets. This might result in either a hazardous main product or dangerous byproducts, with negative effects that can be immediate or long-term. As AI systems become more intelligent, the likelihood of unintended safety issues increases, making these consequences harder to detect and potentially more damaging. Recent studies have highlighted the complexity of unintended outcomes. For instance, AI systems might learn undesired behaviors that are highly rewarded due to misspecified training goals. Similarly, unintended behaviors such as unfaithful explanations during chain-of-thought prompting (\cite{turpin2024language}) or the emergence of deceptive strategies in large language models (\cite{hagendorff2024deception}) underscore the subtleties and risks of unintended consequences.}
These unintended consequences might result in either a hazardous main product or dangerous byproducts, with negative effects that can be immediate or long-term. 
Each scenario necessitates specific detection and prevention strategies to ensure the safe operation of \nc{AI scientists}.

Similarly, each scientific domain in our classification presents distinct risks, each requiring tailored safeguards to mitigate the inherent dangers. 

\vspace{-.2cm}

\vspace{-.2cm} \begin{itemize} \item \jan{\textbf{Natural Science Risks:}} \begin{itemize} \item \jan{\textbf{Chemical Risks} involve the exploitation of agents to synthesize chemical weapons, as well as the creation or release of hazardous substances during autonomous chemical experiments. This category also includes the risks arising from the use of advanced materials, such as nanomaterials, which may have unknown or unpredictable chemical properties.}

    \item \jan{\textbf{Biological Risks} encompass the dangerous modification of pathogens and unethical manipulation of genetic material, potentially leading to unforeseen biohazardous outcomes.}
    
    \item \jan{\textbf{Radiological Risks} involve both immediate operational hazards, such as exposure incidents or containment failures during the automated handling of radioactive materials, and broader security concerns regarding the potential misuse of AI systems in nuclear research.}
    
    \item \jan{\textbf{Physical (Mechanical) Risks} are associated with robotics and automated systems, which could lead to equipment malfunctions or physical harm in laboratory settings. 
    }
\end{itemize}

\vspace{-.2cm}

\item \jan{\textbf{Information Science Risks:} These risks pertain to the misuse, misinterpretation, or leakage of data, which can lead to erroneous conclusions or the unintentional dissemination of sensitive information, such as private patient data or proprietary research. Recent research has demonstrated how LLMs can be exploited to generate malicious medical literature that poisons knowledge graphs, potentially manipulating downstream biomedical applications and compromising the integrity of medical knowledge discovery \cite{yang2024poisoning}. Such risks are pervasive across all scientific domains.}
\end{itemize}

\vspace{-.2cm}

\march{In addition, the impact of \nc{AI scientists} on the external environment spans three distinct domains: the natural environment, human health, and the socioeconomic environment.} 
Risks to the \textit{natural environment} include ecological disruptions and pollution, which may be exacerbated by energy consumption and waste output. 
\march{\textit{Human health} risks encompass damage to both individual and public well-being, such as the negative impact on mental health through the dissemination of inaccurate scientific content.} 
\textit{Socioeconomic} risks involve potential job displacement and unequal access to scientific advancements.
Addressing these risks demands comprehensive frameworks that integrate risk assessment, ethical considerations, and regulatory measures, ensuring alignment with societal and environmental sustainability through multidisciplinary collaboration.

\subsection{Vulnerabilities of \nc{AI Scientists}}
\label{sec:vul}

\begin{figure}[h!]
  \centering  \includegraphics[scale=0.78]{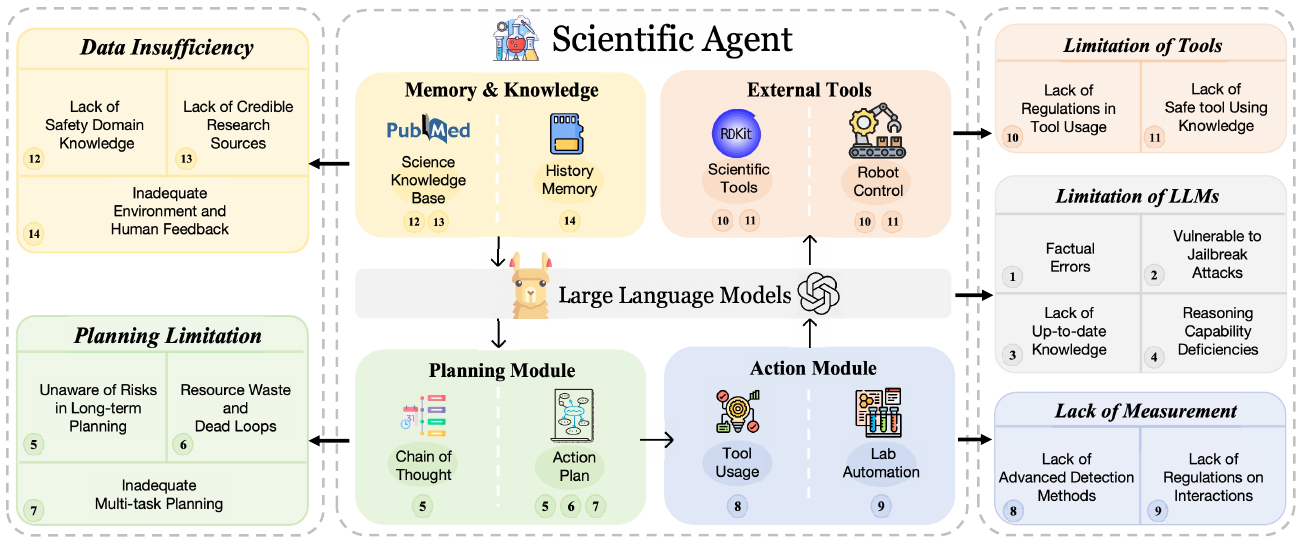}
  \caption{This diagram illustrates the structural framework and potential vulnerabilities of LLM-based \nc{AI scientists}. The agent is organized into five interconnected modules: LLMs, planning, action, external tools, and memory \& knowledge. Each module exhibits unique vulnerabilities. The arrows depict the sequential flow of operations, starting from memory \& knowledge to the use of external tools, underscoring the cyclic and interdependent nature of these modules in the context of scientific discovery and application.}
  \vspace{-.2cm}
  \label{fig:origin}
\end{figure}

LLM-powered agents, including AI scientists, typically encompass five fundamental modules: \emph{LLMs}, \emph{planning}, \emph{action}, \emph{external tools}, and \emph{memory \& knowledge} \cite{park2023generative,wang2023survey}. These modules function in a sequential pipeline: receiving inputs from tasks or users, leveraging memory or knowledge for planning, executing smaller premeditated tasks (often involving scientific domain tools or robotics), and ultimately storing the resulting outcomes or feedback in their memory banks.
Despite the extensive applications, several notable vulnerabilities exist within these modules, giving rise to unique risks and practical challenges (see Figure \ref{fig:origin}). 
In this section, we provide an overview of the high-level concept of each module and summarize the vulnerabilities associated with each.

\subsubsection{LLMs (The Base Models)}
LLMs empower agents with fundamental capabilities. 
However, there are certain risks associated with them:

\textbf{Factual Errors:} LLMs are prone to generating plausible but false information, which is particularly problematic in the scientific domain, where accuracy and trustworthiness are crucial \cite{ji2023survey,bang2023multitask,tian2024opportunities,huang2025survey}.

\textbf{Vulnerable to Jailbreak Attacks:} 
LLMs are susceptible to jailbreak attacks—manipulative prompting techniques that bypass safety measures \cite{zhang2020adversarial,wei2023jailbroken}. \jan{
For instance, while an LLM might refuse a direct request to synthesize illegal substances, it may unknowingly provide instructions when the same request is rephrased using chemical formulas or technical terminology. Similarly, common jailbreak attacks employ creative misdirection through roleplaying scenarios—such as asking the model to help write a fictional play where one character explains dangerous processes to another, or requesting the model to participate in a hypothetical game where harmful information must be shared. 
}

\textbf{Reasoning Capability Deficiencies:} LLMs often struggle with deep logical reasoning and handling complex scientific arguments~\cite{huang-chang-2023-towards,valmeekam2022large,wei2022chain}. Their inability to perform such tasks can result in flawed planning and interaction, as they may resort to using inappropriate tools ~\cite{wornow2023shaky}.

\textbf{Lack of Up-to-Date Knowledge:} LLMs, which are trained on pre-existing datasets, may lack the latest scientific developments, leading to potential misalignments with contemporary scientific knowledge \cite{bommasani2021opportunities}. \jan{Despite the advent of Retrieval-Augmented Generation, challenges remain in sourcing the most recent knowledge. Recent advances in model editing techniques offer promising solutions for efficiently updating LLMs' knowledge in specific domains while preserving performance on other tasks \cite{yao-etal-2023-editing}, though maintaining long-term model relevancy remains an open challenge.}

\subsubsection{Planning Module}

Given a task, the planning module is designed to break down the task into smaller, manageable components.
Nevertheless, the following vulnerabilities exist:

\textbf{Lack of Awareness of Risks in Long-term Planning:}  Agents often struggle to fully comprehend and account for the potential risks associated with their long-term plans of action. This issue arises because LLMs are primarily designed to solve specific tasks rather than to evaluate the long-term consequences of actions with an understanding of potential future impacts~\cite{chui2018real, cave2019bridging}.

\textbf{Resource Waste and Dead Loops:} Agents may engage in ineffective planning processes, leading to resource wastage and becoming stuck in non-productive cycles~\cite{xu2022learning, ruan2023identifying, li2023repetition}. A pertinent example is when an agent is unable to determine whether it can complete a task or continously fails when using a tool it relies on. This uncertainty can cause the agent to repeatedly attempt various strategies repeatedly, unaware that these efforts are unlikely to yield success.

\textbf{Inadequate Multi-task Planning:} Agents often face challenges in handling multi-goal or multi-tool tasks due to their design, which typically optimizes them for single-task performance~\cite{anonymous2024toolllm}. \march{This limitation becomes particularly evident when agents are required to navigate tasks that demand simultaneous attention to diverse objectives or the use of multiple tools in a cohesive manner. The complexity of multi-task planning not only strains the agents' decision-making capabilities but also raises concerns about the reliability and efficiency of their actions in critical scenarios.}

\march{For instance, consider an agent designed to assist in emergency response scenarios, where it must simultaneously coordinate logistics, manage communications, and allocate resources. If the agent is not adept at multi-task planning, it might misallocate resources due to its inability to reconcile the urgency of medical assistance with the need for evacuation efforts. This could result in a delayed response to critical situations, thereby exacerbating the impact of the emergency.}

\subsubsection{Action Module}
\march{Once the task has been decomposed, the action module executes a sequence of actions, specifically by calling tools. }


\textbf{Deficient Oversight in Tool Usage:} A lack of efficient supervision over how agents use tools can lead to potentially harmful situations. For instance, incorrect selection or misuse of tools can trigger hazardous reactions, including explosions. Agents may not be fully aware of the risks associated with the tools they use, \march{as the tools may function as black boxes to the agents}. This is especially true in specialized scientific tasks, \march{where the results of tool usage might be unpredicted and unsafe}.  Thus, it is crucial to enhance safeguards by learning from real-world tool usage.

\textbf{Lack of Regulations on Human-Agent Interactions for Actions:} 
\march{Strengthening regulations on human-agent interactions is crucial as the rising use of agents in scientific discovery highlights the urgent need for ethical guidelines, particularly in sensitive domains like genetics. Despite this, the development of such regulatory frameworks is still at an early stage, as indicated by ~\cite{McConnell2019EthicsValues}. Moreover, the propensity of LLMs to amplify and misinterpret human intentions adds another layer of complexity. Given the decoding mechanisms of LLMs, their limitations in hallucination can lead to the generation of content that presents non-existent counterfactuals, potentially misleading humans.}

 \vspace{-.2cm}

\subsubsection{External Tools}

\oct{During task execution, AI scientists interact with various external software and hardware tools (e.g., robotic arms, chemical analysis software, or molecular design toolkits like RDKit) to accomplish their objectives. While these tools extend the capabilities of AI scientists from planning to physical execution, they also introduce potential risks when misused. For instance, an AI scientist might issue incorrect commands to a robotic arm controller handling chemical substances, potentially leading to hazardous spills or reactions. The challenge lies not in the tools themselves but in the AI scientist's ability to appropriately utilize these specialized external interfaces and anticipate their real-world consequences.}

 \vspace{-.2cm}

\subsubsection{Memory and Knowledge Module}

LLMs' knowledge can become muddled in practice, much like human memory lapses. The memory and knowledge module attempts to mitigate this issue by leveraging external databases for knowledge retrieval and integration. However, several challenges persist:

\textbf{Limitations in Domain-Specific Safety Knowledge:} Agents' knowledge shortfalls in specialties like biotechnology or nuclear engineering can lead to safety-critical reasoning lapses. For instance, an agent for nuclear reactor design might overlook risks like radiation leaks or meltdowns, while an agent involved in compound synthesis may fail to assess toxicity, stability, or environmental impacts \cite{arabi2021artificial}.

\textbf{Limitations in Human Feedback:} Insufficient, uneven, or low-quality human feedback may hinder agents' alignment with human values and scientific objectives. 
Although human feedback plays a crucial role in refining performance and correcting biases, it is often difficult to obtain comprehensively and may not cover all human preferences, especially in complex or ethical scenarios \cite{hagendorff2022methodological}. This underscores the need for improved methods to effectively collect and apply human feedback data.

\textbf{Inadequate Environmental Feedback:} Despite some work on embodied agents~\cite{palm-e}, agents may not receive or correctly interpret environmental feedback, such as the state of the world or the behavior of other agents. This can lead to misinformed decisions that may harm the environment or the agents themselves~\cite {wu2020managing}. For example, an agent trained to manage water resources may not account for rainfall variability, the differing user demands, or the impacts of climate change. 

\textbf{Unreliable Research Sources:} Agents might utilize or train on outdated or unreliable scientific information, leading to the dissemination of incorrect or harmful knowledge. For example, LLMs run the risk of plagiarism \march{of copyrighted content}, content fabrication, or producing false results \cite{jin2023retrieve}.

\tikzstyle{my-box}=[
    rectangle,
    draw=hidden-draw,
    rounded corners,
    text opacity=1,
    minimum height=1.5em,
    minimum width=5em,
    inner sep=2pt,
    align=center,
    fill opacity=.5,
    line width=0.8pt,
]

\tikzstyle{llm}=[my-box, minimum height=1.5em,
    fill=hidden-white!80, text=black, align=center,font=\normalsize,
    inner xsep=2pt,
    inner ysep=4pt,
    line width=0.8pt,
]
\tikzstyle{agent}=[my-box, minimum height=1.5em,
    fill=hidden-yellow!80, text=black, align=center,font=\normalsize,
    inner xsep=2pt,
    inner ysep=4pt,
    line width=0.8pt,
]
\tikzstyle{sci}=[my-box, minimum height=1.5em,
    fill=hidden-orange!80, text=black, align=center,font=\normalsize,
    inner xsep=2pt,
    inner ysep=4pt,
    line width=0.8pt,
]

\tikzstyle{leaf}=[my-box, minimum height=1.5em,
    fill=hidden-pink!80, text=black, align=left, font=\normalsize,
    inner xsep=2pt,
    inner ysep=4pt,
    line width=0.8pt,
]
\begin{figure*}[t!]
    \centering
    \resizebox{\textwidth}{!}{
    \begin{forest}
        forked edges,
        for tree={
            grow=east,
            reversed=true,
            anchor=base west,
            parent anchor=east,
            child anchor=west,
            base=center,
            font=\large,
            rectangle,
            draw=hidden-draw,
            rounded corners,
            align=left,
            text centered,
            minimum width=5em,
            edge+={darkgray, line width=1pt},
            s sep=3pt,
            inner xsep=2pt,
            inner ysep=3pt,
            line width=0.8pt,
            ver/.style={rotate=90, child anchor=north, parent anchor=south, anchor=center},
        },
        where level=1{text width=10em,font=\normalsize,}{},
        where level=2{text width=10em,font=\normalsize,}{},
        where level=3{text width=11em,font=\normalsize,}{},
        where level=4{text width=7em,font=\normalsize,}{},
        [
            LLM Safeguard 
            [
                Safeguarding LLMs, llm
                [
                    Content Evaluation, llm 
                    [
                        Standard, llm 
                        [
                            SafetyBench~\cite{zhang2023safetybench}
                            , leaf, text width=30em
                        ]
                    ]
                    [
                       Alignment-breaking based, llm 
                        [
                            Jailbroken~\cite{wei2023jailbroken}{, }Assert~\cite{mei2023assert}\\
                            BIPIA~\cite{yi2023benchmarking}{, } MasterKey~\cite{deng2023jailbreaker}
                            , leaf, text width=30em
                        ]
                    ]  
                ]
                [
                    Safety Alignment, llm
                    [
                        RLHF, llm
                        [
                            RLHF~\cite{ouyang2022training}{, }\\Safe RLHF~\cite{dai2023safe}
                            , leaf, text width=20em
                        ]
                    ]
                    [
                        Fine-tuning, llm
                        [                            
                            Shadow Alignment~~\cite{yang2023shadow}{, }\\
                            Compromised Fine-tuning~\cite{qi2023fine}{, } \\ 
                            Stay-Tuned LLaMAs~\cite{bianchi2023safety}
                            , leaf, text width=20em
                        ]  
                    ]
                    [
                        Inference, llm
                        [
                            RAIN~\cite{li2023rain}
                            , leaf, text width=20em
                        ]
                    ]
                ]
                [
                    Alignment-breaking \\ Defense, llm
                    [
                        Prompting, llm
                        [
                            Self Defense~\cite{helbling2023llm}{, }RA-LLM~\cite{cao2023defending}{, }\\
                            Goal Prioritization~\cite{zhang2023defending}, leaf, text width=27em
                        ]
                    ]
                    [
                        Parameter Manipulation, llm
                        [
                            Parameter Pruning~\cite{hasan2024pruning}{, }Jatmo~\cite{piet2023jatmo}
                            , leaf, text width=27em
                        ]
                    ]
                ]
            ]
            [
                Safeguarding Agents, agent
                [
                    General Agents, agent
                    [
                        Evaluation, agent
                        [
                            R-Judge~\cite{yuan2024r}{, }AgentMonitor~\cite{naihin2023testing}
                            , leaf, text width=28em
                        ]
                    ]
                    [
                        Risk Detection, agent
                        [
                            Toolemu~\cite{ruan2023identifying} 
                            , leaf, text width=28em
                        ]
                    ]
                ]
                [
                    Scientific Agents, sci
                    [
                        Memory Mechanism, sci
                        [
                            Sciguard~\cite{he2023sciguard}
                            , leaf, text width=28em
                        ]
                    ]
                    [
                        External Tool Using, sci
                        [
                            Chemcrow~\cite{bran2023chemcrow}{, }CLAIRify~\cite{Yoshikawa2023CLAIRify}{, }\\  Coscientist~\cite{Boiko2023Natureagent}
                            , leaf, text width=28em
                        ]
                    ]
                ]
            ]
        ]
    \end{forest}}
\vspace{-.6cm}
\caption{Survey of related work in safeguarding LLMs and agents, among which scientific agents are specifically stated.}
\label{fig:safeguardframework}
  \vspace{-.1cm}
  \vspace{-.1cm}
    \vspace{-.1cm}
\end{figure*}

 \vspace{-.2cm}

\subsection{Recent Work in Safeguarding \nc{AI scientists}}

    \vspace{-.2cm}

\label{sec:safesciagent}

\nc{AI scientists could directly or indirectly produce harmful outputs.} 
\nov{A key concern lies in the gap between syntactic correctness and runtime safety. For example, in programming, both human programmers and AI agents like Copilot can write code that is syntactically correct and appears bug-free, yet may produce unexpected errors or incorrect outputs when deployed. 
Similarly, in chemical experiments, an AI scientist might follow all the correct procedural steps but still inadvertently generate toxic gases or dangerous byproducts during synthesis. 
While human experts can often anticipate and prevent such issues through their experience and knowledge, AI scientists may lack the capability to foresee potential dangerous outcomes.} 
\nc{A survey of recent studies on the risks of LLMs and agents is shown in Figure \ref{fig:safeguardframework} and Table \ref{relatedwork}.}


Coscientist \cite{Boiko2023Natureagent} proposed a chemical agent with access to scientific tools and highlighted the safety risks agents confront, using practical examples to emphasize the need for safety assurance in \nc{AI scientists}. To address these safety concerns, ChemCrow \cite{bran2023chemcrow} introduced a safety tool that reviews user queries to prevent agents from inadvertently creating hazardous chemicals during synthesis in response to malicious commands. 
In addition to filters for user inputs, CLAIRify \cite{Yoshikawa2023CLAIRify} designed specialized safety mechanisms for its chemical agents. 
Furthermore, SciGuard \cite{he2023sciguard} developed a specialized agent for risk control that incorporates long-term memory to enhance safety. 
To evaluate the security of the current models, SciGuard created a benchmark called SciMT-Safety. This benchmark evaluates a model's harmlessness based on its ability to reject malicious queries and gauges its helpfulness based on how effectively it handles benign queries.


\begin{table}[!ht]
\centering
\resizebox{\linewidth}{!}{
\begin{tabular}{|>{\raggedright\arraybackslash}p{2cm}|>{\raggedright\arraybackslash}p{10cm}|>{\raggedright\arraybackslash}p{3cm}|}
\hline
\textbf{Type of Safety Risk} & \textbf{LLMs} & \textbf{AI Scientists (agents)} \\
\hline
\textbf{Content Safety Risks} &
\textbf{Risks Identified:} Issues such as offensiveness, unfairness, illegal activities, and ethical concerns \cite{zhang2023safetybench,zhiheng2023safety}. \newline
\textbf{Evaluation Methods:} SafetyBench with multiple-choice questions covering seven categories of safety risks \cite{zhang2023safetybench}. \newline
\textbf{Alignment Methods:} Reinforcement learning from human feedback (RLHF) \cite{ouyang2022training,casper2023open}. Safe RLHF, decoupling helpfulness and harmlessness \cite{dai2023safe}. Self-evaluation and training-free alignment via RAIN \cite{li2023rain}. \newline
\textbf{Fine-tuning Safety:} Adversarial examples and benign data can inadvertently compromise model safety during fine-tuning \cite{qi2023fine, yang2023shadow}. Reassuringly, extra safety examples can improve this concern, an excess may hinder it \cite{bianchi2023safety}.
&
\textbf{Tool Interaction Risks:} Identifying risks of agents with an emulator \cite{ruan2023identifying}. \\
\hline
\textbf{Jailbreak Vulnerabilities} &
\textbf{Alignment-Breaking Attacks:} Evaluated under jailbreaking conditions \cite{wei2023jailbroken, deng2023jailbreaker, mei2023assert, yi2023benchmarking}. \newline
\textbf{Defenses:} Prompt techniques (self-examination) \cite{helbling2023llm, zhang2023defending, cao2023defending}, parameter pruning \cite{hasan2024pruning}, fine-tuning \cite{piet2023jatmo}.
&
\textbf{Evaluation of Risk Awareness:} Techniques like AgentMonitor \cite{naihin2023testing} and R-Judge \cite{yuan2024r}. \\
\hline
\end{tabular}}
\caption{\nc{Summary of LLMs and AI scientists (agents) safety concerns and solutions.} \nov{Here, LLMs refer to base language models that primarily process and generate text, while AI scientists are autonomous systems that combine LLMs with the ability to use external tools (e.g., laboratory equipment, scientific software) and take actions in the physical world. For example, while an LLM might generate text describing a chemical reaction, an AI scientist could execute that reaction using robotic equipment.}}
\label{relatedwork}
\end{table}

\vspace{-.2cm}

\subsection{Current Limitations}

Since \nc{AI scientists} confront ubiquitous risks, effective safety mechanisms should consider user inputs, agent actions, and environmental consequences. However, current efforts remain incomplete.

\textbf{(1) Lack of safety constraints on action space.}
Most work on safeguarding AI scientists demand the use of external tools ~\cite{he2023sciguard}. However, the limited capabilities of agents can lead to the unintentional misuse of tools and harmful outcomes, which can be more severe when misled by adversaries. \oct{A fundamental solution is to constrain the input domain of possible actions. Leading agent frameworks~\cite{autogpt2023,ruan2023identifying} demonstrate this by predefining a fixed and finite action space to balance safety and functionality.} For example, AutoGPT limits a code agent's file system access to `read\_file' operations only, preventing potentially dangerous `write\_file' operations. \oct{Such domain constraints on tool functions can be systematically applied when developing AI scientists to ensure safer operation.}

\textbf{(2) Lack of specialized models for risk control.} 
\oct{Apart from SciGuard~\cite{he2023sciguard}, specialized safety mechanisms for AI scientists are largely lacking. Current approaches mainly rely on input filtering to prevent harmful commands or LLM-based monitoring~\cite{ruan2023identifying, naihin2023testing, yuan2024r} to screen agent behaviors during execution. However, more proactive approaches, such as adversarial models explicitly trained to identify potential exploits in AI scientists, are needed, similar to GAN-style security testing. These specialized safety measures are particularly crucial given the high-stakes nature of scientific experiments compared to general web or software tasks.}

\oct{\textbf{(3) Lack of domain-specific expert knowledge.} Compared with general-purpose agents that handle web browsing~\cite{yao2022webshop} or basic tool usage~\cite{schick2023toolformer}, AI scientists require sophisticated domain expertise. For example, synthesizing small molecules demands deep biochemistry knowledge to understand molecular properties and reaction mechanisms. Such expertise is critical for two aspects of safety: (1) enabling proper experimental planning and tool usage to prevent accidents, and (2) recognizing potential hazards in advance. For instance, an agent with chemistry expertise would understand that certain chemical combinations can trigger dangerous exothermic reactions and avoid such combinations.}

\textbf{(4) Ineffective evaluations on the safety of \nc{AI scientists}.} To date, benchmarks evaluating safety in the scientific realm, such as SciMT-safety~\cite{he2023sciguard}, only consider the harmlessness of models by examining their ability to deny malicious requests. Considering the multifaceted issues mentioned above, safeguarding \nc{AI scientists} demands additional benchmarks focused on comprehensive risk scopes (Section \ref{sec:chap2problem}) and various agent vulnerabilities (Section \ref{sec:vul}).

\vspace{-.2cm}

\section{Proposition}
\textit{It has become increasingly evident that \nov{developers} must prioritize risk control over autonomous capabilities.}
While autonomy is an admirable goal and significant for enhancing productivity across various scientific disciplines, it cannot be pursued at the expense of generating serious risks and vulnerabilities. 
Consequently, we must balance autonomy with security and employ comprehensive strategies to ensure the safe deployment and use of \nc{AI scientists}.

Moreover, the emphasis should shift from output safety to behavioral safety, which signifies a comprehensive approach that evaluates not only the accuracy of the agent's output but also the actions and decisions it takes. 
Behavioral safety is critical in the scientific domain, as the same action in different contexts can lead to vastly different consequences, some of which may be detrimental. 
Here, we propose fostering a triadic relationship involving humans, machines, and the environment. 
This framework recognizes the importance of robust and dynamic environmental feedback, in addition to human feedback.

\jan{To address current limitations in safety requirements and domain expertise, we propose a dual-pronged interim strategy that combines enhanced human supervision with conservative operational constraints. First, we recommend implementing heightened expert oversight in domains where autonomous safety measures are still evolving, ensuring continuous monitoring and validation of AI system behaviors. Second, we advocate restricting autonomous operations to well-characterized, lower-risk scenarios where safety parameters and operational boundaries have been thoroughly validated through empirical testing and expert review.
}
 
\subsection{Agent Alignment and Safety Evaluation} 
\subsubsection{Agent Alignment}

\textbf{Improving LLM Alignment:} 
The foundation of AI scientist safety lies in better-aligned LLMs—ensuring they generate responses that adhere to safety guidelines and legal requirements. Current alignment efforts focus on several complementary approaches: filtering harmful or illegal content through careful data curation, applying Constitutional AI principles \cite{bai2022constitutional}, \jan{and using targeted knowledge editing techniques to detoxify model behaviors \cite{wang-etal-2024-detoxifying}.}

\textbf{Towards Agent-level Alignment:} 
\oct{Agent alignment, however, presents a unique challenge: controlling sequences of actions that may be individually benign but potentially harmful in specific contexts. While LLM alignment can be achieved through output filtering, agent alignment requires understanding and replicating human expert workflows. For instance, in biological research, an agent needs to learn not just what to do, but how expert researchers systematically investigate genetic variants—consulting literature, analyzing similar variants, and understanding gene interactions. This kind of sequential decision-making cannot be learned through simple prompting or output filtering. Instead, it requires:
(1) comprehensive datasets of human expert workflows, capturing step-by-step research methodologies;
(2) domain experts providing feedback on action sequences, similar to how autonomous driving systems learn from real-world driving data; and
(3) reward models that evaluate not just individual actions but entire research strategies.
The key challenge is that, while we have abundant data on what researchers write (e.g., papers, answers), we lack structured data on how they conduct research—their sequence of actions, tool usage, and decision-making processes.
}

\subsubsection{Safety Evaluation}
\textbf{Red Teaming:} 
Identifying potential vulnerabilities that may cause hazardous activities to users and the environment is essential for evaluating agent safety.
Red-teaming\cite{feffer2024redteaming}, \emph{i.e.}, adversarially probing LLMs for harmful outputs, has been widely used in the development of general LLMs.  For example, jailbreaks that challenge model safety are used in red-teaming evaluations and have been specifically noted as alignment-breaking techniques in Table \ref{relatedwork}. Furthermore, red-teaming datasets can be utilized to train LLMs for harm reduction and alignment reinforcement. However, specialized red-teaming for \nc{AI scientists} is absent. Considering the severe risks in the scientific domain (Section \ref{sec:chap2problem}), we advocate for red-teaming against \nc{AI scientists}.
\jan{The criteria for effective red-teaming of AI scientists include:
(1) Domain-specificity: Testing scenarios must reflect realistic scientific workflows and domain-specific safety concerns;
(2) Complexity gradients: Scenarios should progress from simple protocol deviations to complex multi-step safety violations;
(3) Cross-domain interactions: Tests should examine how safety measures in one domain affect operations in others.
Our initial validation tests on chemical synthesis agents demonstrate the effectiveness of these criteria, though broader testing across different scientific domains is ongoing. Red-teaming for AI scientists differs from general LLM testing in several key aspects:
(1) Physical safety implications: Tests must account for real-world consequences beyond text generation;
(2) Domain expertise requirements: Red team members need both security expertise and domain-specific knowledge;
(3) Tool interaction complexity: Tests must cover both language model responses and tool usage patterns.
}

\textbf{Benchmarking:} 
To address the various risks stated in Section \ref{sec:chap2problem}, comprehensive benchmarks should cover a wider range of risk categories and provide a more thorough coverage across domains. 
To address vulnerabilities stated in Section \ref{sec:vul}, effective benchmarks should focus on various dimensions such as tool usage \cite{huang2024planning}, risk awareness \cite{naihin2023testing, yuan2024r}, and resistance to red-teaming \cite{deng2023jailbreaker, mei2023assert, yi2023benchmarking}.

\jan{
\textbf{Task Alignment:} Our framework implements a graduated autonomy approach to address the challenge of maintaining agent performance while ensuring safety. The agent begins with restricted operations in well-defined, lower-risk tasks and gradually expands its operational scope as safety metrics are met. This is complemented by continuous monitoring systems that evaluate both task performance and safety compliance. When performance metrics indicate degradation due to safety constraints, the system triggers a human expert review to optimize the balance between safety and functionality. This approach allows for dynamic adjustment of safety parameters based on task complexity and risk level, rather than applying uniform restrictions across all operations.}

\vspace{-.2cm}\vspace{-.3cm}

\subsection{Human Regulation}
In addition to steering already-capable models, it is also important to impose certain regulations on the developers and users of these highly capable models.

   \vspace{-.3cm}

\vspace{-.3cm}

\subsubsection{Developer Regulation}
The primary goal of developer regulation is to ensure \nc{AI scientists} are created and maintained in a safe, ethical, and responsible manner. 
\jan{Similar to how automobile manufacturers must meet safety standards and certification requirements before being authorized to produce vehicles, developers should be required to obtain certification before being authorized to develop AI scientists. }

\vspace{-.3cm}
First, developers of \nc{AI scientists} should adhere to \nov{an internationally recognized} code of ethics. This includes mandatory training in ethical AI development, with an emphasis on understanding the potential societal impacts of their creations \nov{across global contexts}. 
\nov{Second, we need a practical framework for safety and ethical compliance checks that can work across jurisdictions. 
This could combine international standards, regional certification bodies, automated testing tools, and peer review mechanisms, though enforcing such oversight globally remains challenging.}

Furthermore, developers should implement robust security measures to prevent unauthorized access and misuse. This includes ensuring data privacy, securing communication channels, and safeguarding against cyber threats. \nov{The development life cycle should incorporate regular security assessments conducted by both internal teams and independent third-party auditors, although establishing consistent international oversight remains challenging.}
Lastly, there should be transparency in the development process. Developers must maintain detailed logs of their development activities, algorithms used, and decision-making processes. These records should be accessible for audits and reviews, ensuring accountability and facilitating continuous improvement.

\subsubsection{User Regulation}
\jan{Regulating the use of autonomous agents in research is crucial. First, potential users should obtain a license to access AI scientists, analogous to how drivers must be licensed before operating vehicles.}
To acquire this license, users should be required to undergo relevant training and pass a knowledge evaluation on the responsible use of \nc{AI scientists}.
\nov{Usage monitoring should balance safety oversight with laboratory privacy, focusing on critical safety incidents and anonymized usage patterns while respecting institutional autonomy and intellectual property rights.}

\nov{Similar to clinical studies requiring Institutional Review Board (IRB) approval, autonomous scientific research needs institutional oversight. However, rather than relying on researcher self-disclosure, specialized committees with AI safety expertise should provide standardized risk assessment protocols and evaluations.}

 \jan{Our framework implements a layered oversight approach:
(1) Institution-Level Controls: Primary oversight resides with institutional review boards specifically trained in AI safety protocols, allowing organizations to maintain control over their research processes while ensuring compliance.
(2) Privacy-Preserving Auditing: External safety monitoring focuses on aggregated metrics and anonymized usage patterns rather than granular research details. This approach enables effective safety oversight while protecting sensitive intellectual property and research data.
(3) Tiered Reporting Structure: A graduated reporting system where only critical safety incidents require detailed external review, with clear guidelines protecting proprietary information and research confidentiality.
However, more thorough safety checks inevitably increase response latency. This time-complexity trade-off means that achieving higher safety standards often comes at the cost of decreased operational speed, potentially limiting real-time applications. 
}

\vspace{-.3cm}

\subsection{Agent Regulation and Environmental Feedback}

Understanding and interpreting environmental feedback is critical for \nc{AI scientists} to operate safely. Such feedback includes various factors, such as the physical world, societal laws, and developments within the scientific system.

\textbf{Simulated Environment for Result Anticipation:} \nc{AI scientists} can significantly benefit from training and operating within simulated environments designed specifically to mimic real-world conditions and outcomes. This process allows the model to gauge the potential implications of certain actions or sequences of actions without causing real harm. For example, in a simulated biology lab, an autonomous agent can experiment and learn that improper handling of biohazardous material can lead to environmental contamination. Through trials within the simulation, the model can understand that specific actions or procedural deviations may lead to dangerous situations, helping to establish a safety-first operating principle.

\jan{Our simulated environments are evaluated using:
(1) Physical fidelity metrics comparing simulation outputs with real-world experimental results across key parameters;
(2) Process fidelity metrics measuring the accuracy of simulated workflow sequences against recorded laboratory procedures;
(3) Error propagation analysis to understand how simulation uncertainties affect decision outcomes.
The environments undergo continuous calibration using real-world feedback, with particular attention to edge cases and failure modes identified during actual laboratory operations.
}

\textbf{Agent Regulation:} Agent regulation may focus on the symbolic control of autonomous agents~\cite{hong2023metagpt} and multi-agent or human-agent interaction scenarios. A specialized design, such as a ``safety check'' standard operating procedure, could be applied to control when and how agents utilize scientific tools that could be exploited for malicious intents or result in unintended consequences. \jan{Specifically, to mitigate the risk of unintended consequences, agents could be programmed to incorporate dynamic safety checks that assess not only the direct effects of their actions but also potential secondary or indirect impacts. Additionally, the implementation of a consequence-aware regulation system could require agents to simulate and evaluate the long-term consequences of their actions before execution.} Another possible solution is to require autonomous agents to obtain approval from a committee consisting of human experts before each query involving critical tools and APIs that may lead to potential safety concerns.

\jan{\textbf{Real-time Decision Making:} Our framework implements a multi-level decision validation system:
(1) A fast-response layer for immediate safety-critical decisions using pre-validated action templates;
(2) A medium-latency layer for complex decisions requiring rapid but non-immediate responses, incorporating real-time environmental feedback;
(3) A deliberative layer for decisions with longer-term implications, allowing for a more comprehensive risk assessment.
This hierarchical approach enables the system to balance response speed with safety considerations while maintaining operational efficiency.
}

\textbf{Critic Models:} Beyond standard safety checks, specialized oversight models can play crucial roles in safety verification. Critic models can serve as additional layers that assess and refine outputs. By identifying potential errors, biases, or harmful recommendations, critic models contribute significantly to reducing risks associated with the AI's operation \cite{xu2021machine,mohseni2022taxonomy}.
\oct{Additionally, adversarial models, similar to GANs, can be specifically trained to identify potential exploits and vulnerabilities.}
   
\textbf{Tuning Agents with Action Data:} Unlike the setup for LLM alignment, where the aim is to train the LLM or directly impose an operational procedure on an agent, using annotated data that reflect potential risks of certain actions can enhance agents' anticipation of harmful consequences. By leveraging extensive annotations made by experts—such as marking actions and their results during laboratory work—we can continue to fine-tune agents. For example, a chemical study agent would understand that certain mixes can lead to harmful reactions. Additionally, training should incorporate mechanisms that limit agents' access to dangerous tools or substances, relying on annotated data or simulated environmental feedback. In biochemistry or chemical labs, agents could learn to avoid interactions that may lead to biohazard contamination or hazardous reactions.
\jan{To address the gaps in sequential decision-making data, our framework employs three complementary strategies:
(1) Hybrid data collection combining direct expert observation with automated workflow logging.
(2) Synthetic data generation using validated expert-designed templates: This approach creates diverse simulated interaction scenarios specifically designed to test the AI scientist's decision-making capabilities across a spectrum of challenging conditions. Similar to how automobile manufacturers test vehicles on specially designed courses with various obstacles, steep gradients, and difficult terrain before real-world deployment, we can systematically generate synthetic interaction scenarios that don't necessarily correspond to specific real-world use cases but effectively stress-test the system's safety boundaries. These synthetic scenarios would include adversarial prompts, edge cases, intentionally ambiguous instructions, and complex multi-step tasks with hidden safety implications. And
(3) Active learning approaches where the system identifies knowledge gaps and requests specific expert demonstrations.
Additionally, we could implement a confidence-based execution system where actions with insufficient supporting data require explicit expert validation before execution.
}

\section*{Acknowledgement}

Xiangru Tang and Mark Gerstein are supported by Schmidt Futures. Qiao Jin and Zhiyong Lu are supported by the NIH Intramural Research Program, National Library of Medicine.


\section*{Competing Interests}
The authors declare no competing interests.

\section*{Author Contributions Statements}

XT and QJ co-initiated the project, with XT leading the writing of the manuscript and contributed to the theoretical framework. XT, QJ, KZ, TY, and YCZ contributed to writing specific chapters. WZ and YLZ contributed to the risk analysis framework. MQ and JT provided expertise on drug design applications. ZZ contributed to the safety alignment approaches. AC provided guidance on natural language processing. DG contributed expertise on ethical considerations and regulatory frameworks. ZL provided domain expertise on biomedical applications. MG supervised the overall project and provided strategic guidance. XT was responsible for revising and refining the entire manuscript. All authors provided valuable suggestions and feedback, contributed to the interpretation of results, and approved the final version of the paper.



\normalem

\end{document}